\documentclass[a4paper,12pt]{article}

\usepackage{graphicx}

\begin{document}

\begin{center}
{\large\bf Correlation observables in $\Upsilon D$ pair production at
the LHC within the parton Reggeization approach}

{A.V. Karpishkov$^{1}$, M.A. Nefedov $^{1,2}$, V.A. Saleev$^{1,\P}$}

$^1${Samara National Research University, Moskovskoe Shosse 34,
846064, Samara, Russia}\\

$^2${II. Institut f\"ur Theoretische Physik, Universit\"at Hamburg, Luruper Chaussee 149, 22761 Hamburg, Germany }

$^\P${E-mail: saleev@samsu.ru}
\end{center}

\centerline{\bf Abstract} We study angular correlations in
associated hadroproduction of $\Upsilon(1S)$ with the $D^{\pm}$ and
$D^0$-mesons at the LHC in the Leading Order of the parton
Reggeization approach. Hadronization of $b\bar{b}$-pair to
$\Upsilon(1S)$ is described within the NRQCD-factorization
framework. Production of $D-$mesons is described in the
fragmentation model with scale-dependent fragmentaion functions. We have found good agreement with
LHCb data for various differential distributions, except for the case of spectra on azimuthal angle
differences at the small $\triangle\varphi$ values. The total cross-section in our Single Parton Scattering model,
calculated under conservative assumptions, accounts for almost one half of observed cross-section, thus dramatically shrinking
the room for Double Parton Scattering mechanism. \\



\section{Introduction}

Theoretical and experimental study of angular and momentum
correlations in pair production of hadrons or jets in high energy
hadronic collisions is very important task for various reasons.
First, this is a challenging test of our understanding of
higher-order corrections in quantum chromodynamics (QCD). In general
it is a nontrivial task to provide reliable predictions for such
multiscale and correlational observables, based on the conventional
Collinear Parton Model (CPM) of QCD. Second, if heavy quarkonia are
involved, various models of hadronization of heavy quark pair, such
as NRQCD factorization formalism~\cite{NRQCD}, can be tested. And
last but not least, correlational observables in pair-production of
jets, photons and heavy mesons has become a primary tool in
experimental searches of manifestations of Double Parton
Scattering(DPS) mechanism in proton-proton collisions at high
energies, see e.g.~\cite{LHCb_ups_D, LHCb_DD}.

In the present contribution we discuss associated production of
bottomonia and open charm hadrons in forward rapidity region in $pp$
collisions at the $\sqrt{S} = 7$ and $8$ TeV which has been observed
by the LHCb Collaboration~\cite{LHCb_ups_D}. Comparison of
theoretical predictions obtained in leading order (LO) of CPM with
the measured cross-sections and the differential distributions
pointed towards DPS as the main production mechanism. However,
recently~\cite{PRD2016DD}, we have found that total cross sections
and different spectra of same-sign ($DD$) pairs at LHCb
\cite{LHCb_DD} can be described by the Single Parton Scattering
(SPS) mechanism, if higher-order QCD corrections are approximately
taken into account using the parton Reggeization approach
(PRA)~\cite{NSS2013,NKS2017}. The gauge-invariant amplitude of gluon
pair production in the fusion of two off-shell(Reggeized) gluons
together with scale-dependent $g\to D$ fragmentation functions where
key ingredients of calculations in Ref.~\cite{PRD2016DD}. In the
present note, we present the study of production of
$\Upsilon(1S)D^0$ and $\Upsilon(1S)D^+$ pairs in the combined
approach based on PRA, nonrelativistic QCD (NRQCD)\cite{NRQCD}
factorization and fragmentation model.

\section{Parton Reggeization Approach}
The brief description of LO approximation of PRA is presented below.
More details can be found in \cite{NKS2017}, the development of PRA
in the NLO approximation is further discussed in \cite{PRANLO1}. The
main ingredients of PRA are factorization formula, unintegrated
parton distribution functions (unPDF's) and gauge-invariant
amplitudes with off-shell initial-state partons, derived using the
Lipatov's Effective Field Theory (EFT) of Reggeized
gluons~\cite{Lipatov95} and Reggeized
quarks~\cite{LipatovVyazovsky}.

Factorization formula of PRA in LO approximation for the process
$p+p\to {\cal Y}+ X$, can be obtained from factorization formula of
the collinear parton model (CPM) for the auxiliary hard subprocess
$g+g\to g + {\cal Y}+ g$. In the Ref.~\cite{NKS2017} the modified
Multi-Regge Kinematics (mMRK) approximation for the auxiliary
amplitude is constructed, which correctly reproduces the Multi-Regge
and collinear limits of corresponding QCD amplitude. This
mMRK-amplitude has $t$-channel factorized form, which allows one to
rewrite the cross-section of auxiliary subprocess in a
$k_T$-factorized form:
  \begin{eqnarray}
  d\sigma &=& \int\limits_0^1 \frac{dx_1}{x_1} \int \frac{d^2{\bf q}_{T1}}{\pi} \tilde{\Phi}_g(x_1,t_1,\mu^2)
\int\limits_0^1 \frac{dx_2}{x_2} \int \frac{d^2{\bf q}_{T2}}{\pi}
\tilde{\Phi}_g(x_2,t_2,\mu^2)\cdot d\hat{\sigma}_{\rm PRA},
\label{eqI:kT_fact}
  \end{eqnarray}
where $t_{1,2}=-{\bf q}_{T1,2}^2$, the partonic cross-section
$\hat\sigma_{\rm PRA}$ in PRA is determined by squared PRA
amplitude, $\overline{|{\cal A}_{PRA}|^2}$. Despite the fact that
four-momenta ($q_{1,2}$) of partons in the initial state of ${\cal
A}_{PRA}$ are off-shell ($q_{1,2}^2=-t_{1,2}<0$), the PRA
hard-scattering amplitude is {\it gauge-invariant} because the
initial-state off-shell gluons are treated as Reggeized gluons ($R$)
in a sence of gauge-invariant EFT for QCD processes in Multi-Regge
Kinematics(MRK), introduced by L.N. Lipatov in~\cite{Lipatov95}. The
Feynman rules of this EFT are written down in
Ref.~\cite{AntonovLipatov}.


 The tree-level ``unintegrated PDFs'' (unPDFs) $\tilde{\Phi}_g(x_{1,2},t_{1,2},\mu^2)$ in Eq. (\ref{eqI:kT_fact}) are equal to the convolution
of the collinear PDF $f_g(x,\mu^2)$ and DGLAP splitting function
$P_{gg}(z)$ with the factor $1/t_{1,2}$. Consequently, the
cross-section (\ref{eqI:kT_fact}) with such ``unPDFs'' contains the
collinear divergence at $t_{1,2}\to 0$ and infrared (IR) divergence
at $z_{1,2}\to 1$. To regularize the latter, we observe, that the
mMRK expression gives a reasonable approximation for the exact
matrix element only in the rapidity-ordered part of the phase-space
$y_{g_1}>y_{\cal Y}>y_{g_2}$. From this requirement, the following
cutoff on $z_{1,2}$ can be derived: $z_{1,2}<
1-\Delta_{KMR}(t_{1,2},\mu^2),$
  where $\Delta_{KMR}(t,\mu^2)=\sqrt{t}/(\sqrt{\mu^2}+\sqrt{t})$ is the KMR cutoff function~\cite{KMR}, and we have
 taken into account that $\mu^2\sim M_{T{\cal Y}}^2$. The collinear
singularity is regularized by the Sudakov formfactor:
  \begin{equation}
  T_i(t,\mu^2)=\exp\left[ - \int\limits_t^{\mu^2} \frac{dt'}{t'} \frac{\alpha_s(t')}{2\pi}
\sum\limits_{j=q,\bar{q},g} \int\limits_0^{1} dz\ z\cdot P_{ji}(z)
\theta\left(1-\Delta_{KMR}(t',\mu^2) - z\right)  \right],
\label{eq:Sudakov}
  \end{equation}
  which resums doubly-logarithmic corrections $\sim\log^2 (t/\mu^2)$ in the LLA.

The final form of our unPDF in PRA is:
  \begin{equation}
  \Phi_i(x,t,\mu^2) = \frac{T_i(t,\mu^2)}{t} \frac{\alpha_s(t)}{2\pi} \sum_{j=q,\bar{q},g}
   \int\limits_x^{1} dz\ P_{ij}(z)\cdot \frac{x}{z}f_{j}\left(\frac{x}{z},t \right)\cdot \theta\left(1-\Delta_{KMR}(t,\mu^2)-z \right), \label{eqI:KMR}
  \end{equation}
which coincides with Kimber, Martin and Ryskin (KMR)
unPDF~\cite{KMR}. The KMR unPDF is actively used in the
phenomenological studies employing $k_T$-factorization, but to our
knowledge, the derivation, presented in~\cite{NKS2017} is the first
systematic attempt to clarify it's relationships with MRK limit of
the QCD amplitudes.

In contrast to most of studies in the $k_T$-factorization, the gauge-invariant matrix elements with off-shell initial-state partons (Reggeized
quarks and gluons) from Lipatov's EFT~\cite{Lipatov95,
LipatovVyazovsky} allow one to study arbitrary processes
involving non-Abelian structure of QCD without violation of
Slavnov-Taylor identities due to the nonzero virtuality of initial-state
partons. This approach, together with KMR unPDF gives stable and
consistent results in a wide range of phenomenological applications,
which include the description of the angular correlations of
dijets~\cite{NSS2013}, $b$-jets~\cite{SSbb}, charmed~\cite{PLB2016}
and bottom-flavored~\cite{NKS2017} mesons, as
well as some other examples.

\section{Model for $\Upsilon D$ pair production}

Considering the inclusive production of $\Upsilon(1S)+D$ meson
pairs, one should take into account both direct and feeddown
production of $\Upsilon(1S)$. In the direct case, and taking into
account the possibility of $g\to D$ and $c\to D$ fragmentation, one
has the following LO subprocesses in PRA:
\begin{eqnarray}
R + R &\to& \Upsilon(1S)+ g(\to D), \label{RRg}\\
R + R &\to& \Upsilon(1S)+c(\to D)+\bar c \label{RRcc}.
\end{eqnarray}
The feed-down contribution includes decays of higher
$S-$wave and $P-$wave states:
\begin{eqnarray}
R + R &\to& \Upsilon(2S,3S)+ g(\to D),  \qquad R + R \to  \Upsilon(2S,3S)+c(\to D)+\bar c,\label{Sg}\\
R + R &\to& \chi_b(2P,1P)+ g(\to D),  \qquad R + R \to \chi_b(2P,1P)+c(\to D)+\bar
c\label{Pg}.
\end{eqnarray}
According to NRQCD factorization formalism \cite{NRQCD} final heavy quarkonium can be
produced via color-singlet and color-octet intermediate states. We
will use the set of color-singlet and color-octet nonperturbative
matrix elements, which has been obtained in Ref.~\cite{NSSUpsilon}
from the fit of inclusive $p_T-$spectra of $\Upsilon(nS)$,
measured at the the LHC.

To describe the $D$-meson production we use the fragmentation model
in which the transition of gluon or $c-$quark to the
$D-$meson is described by corresponding fragmentation function (FF)
$D_{c,g}(z,\mu^2)$.  We use universal scale-dependent $c$-quark and
gluon fragmentation functions fitted to $e^+e^-$-annihilation data
from CERN LEP1, in the Ref. \cite{KKSS}. The same FFs has been used earlier in the
description of  $DD-$pair production, measured by the LHCb
collaboration \cite{PRD2016DD}.

\section{Results and discussion}

\begin{figure}[]
\includegraphics[width=0.5\textwidth, angle=-90,origin=c, clip=]{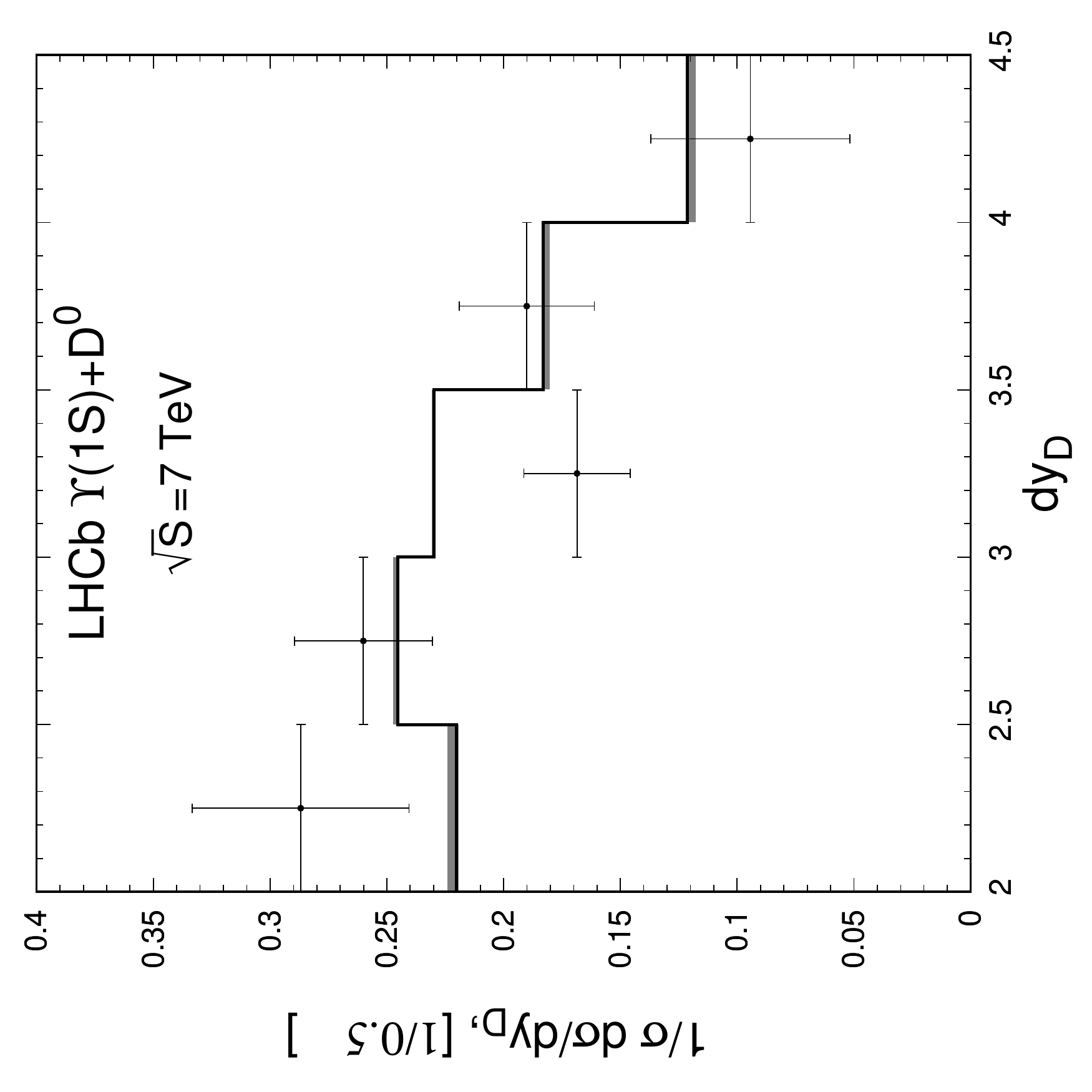}\includegraphics[width=0.5\textwidth, angle=-90,origin=c, clip=]{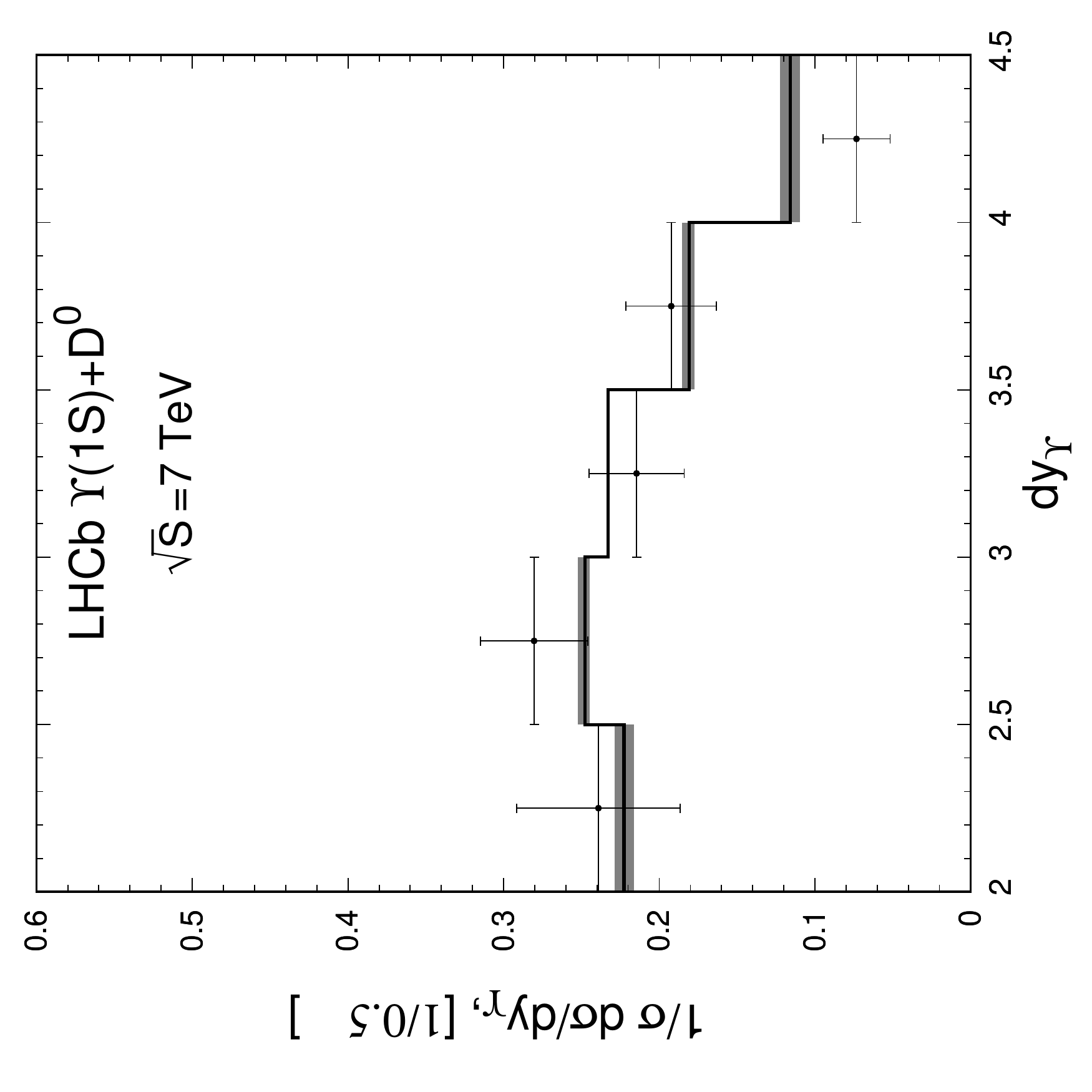}
\includegraphics[width=0.5\textwidth, angle=-90,origin=c, clip=]{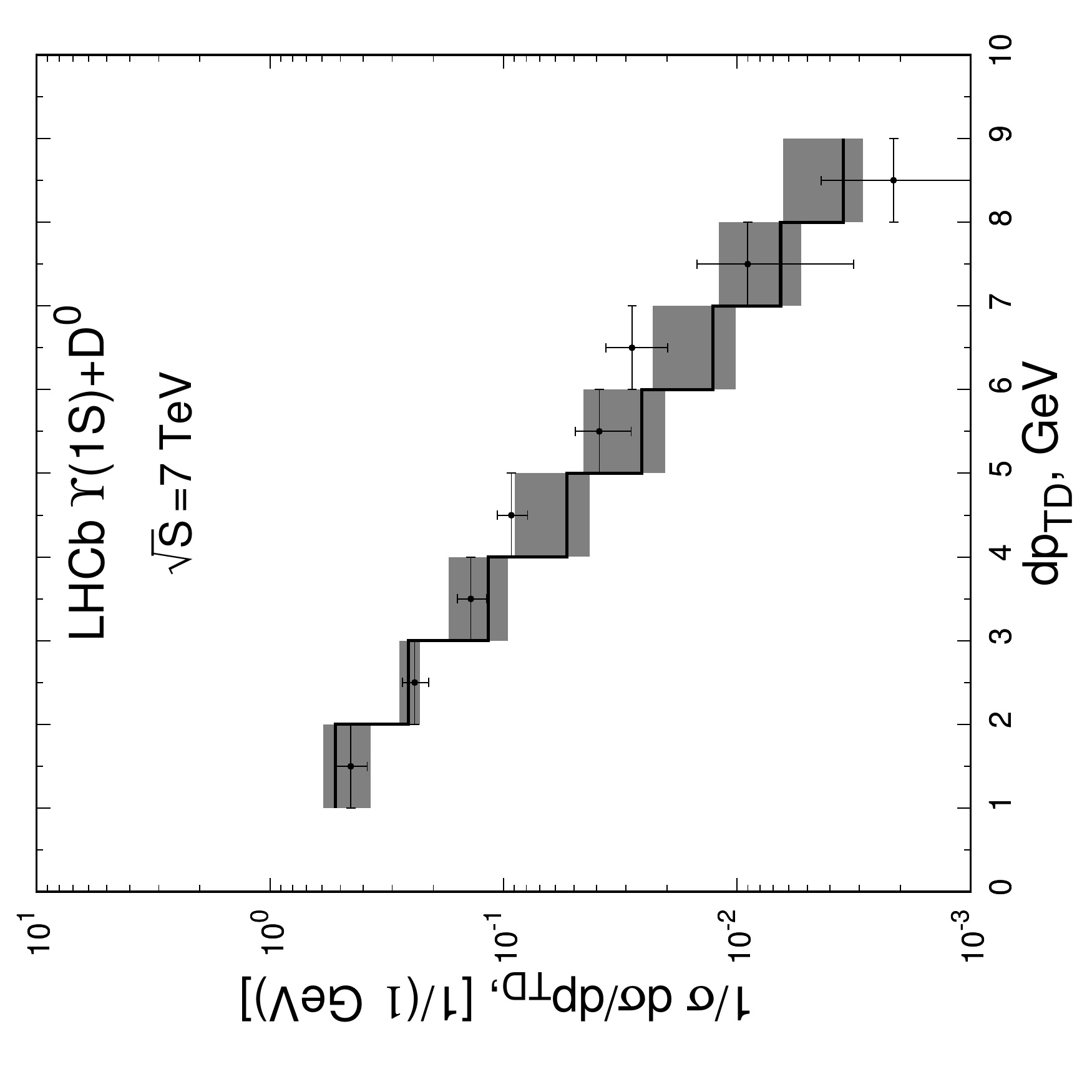}\includegraphics[width=0.5\textwidth, angle=-90,origin=c, clip=]{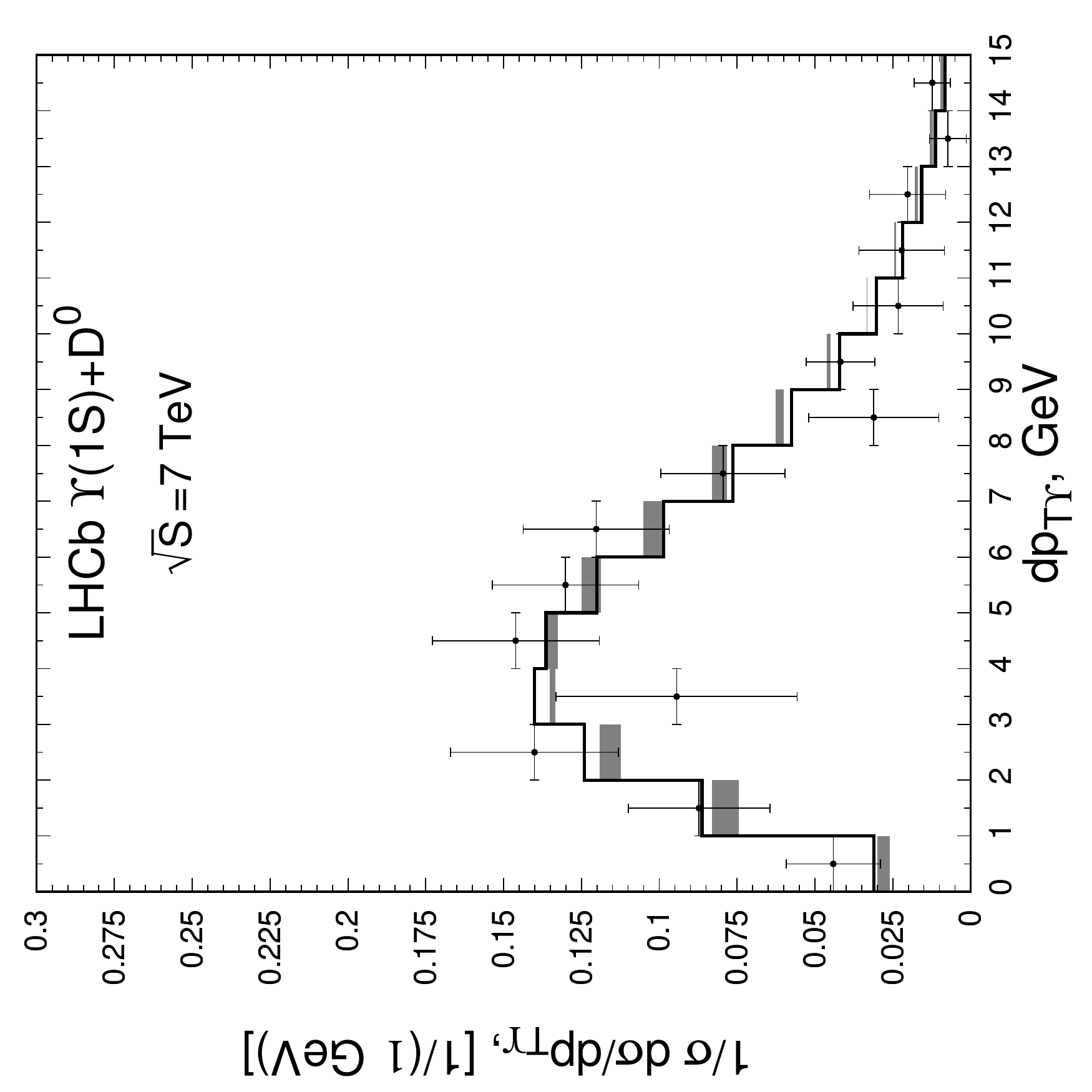}
 \caption{Transverse momentum and rapidity spectra of  $\Upsilon(1S)$ and $D^0$ in the associated production.
 The data from LHCb at the $\sqrt S=7$ TeV, $2.0<y_{\Upsilon(D)}<4.5$, $0<p_{T\Upsilon}< 15$ GeV, and $1<p_{TD}< 20$ GeV.}
 \label{fig-2}
\end{figure}

Here we discuss our results obtained for prompt
$\Upsilon(1S)D^{0,+}$ pair production in $pp$-collisions at
$\sqrt{S}=7$ TeV. Due to the lack of space, in the figures below,
normalized spectra are plotted only for $\Upsilon(1S)D^{0}$
production. The results for $\Upsilon(1S)D^{+}$ are similar. In the
Fig. \ref{fig-2} we demonstrate, that transverse momentum and
rapidity spectra of $\Upsilon(1S)$ and $D^{0}$ are described well in
the LO of PRA. We set the renormalization and factorization scales
to
$\mu_R=\mu_F=\frac{\xi}{2}\left(\sqrt{M_{\Upsilon}^2+p_{T\Upsilon}^2}+\sqrt{M_{D}^2+p_{TD}^2}\right)$
where $\xi = 1$ for the central lines of our predictions, and we
vary $1/2 < \xi < 2$ to estimate the scale-uncertainty of our
prediction, which is shown in the figures by the gray band.

\begin{figure}[]
 \includegraphics[width=0.5\textwidth, angle=-90,origin=c, clip=]{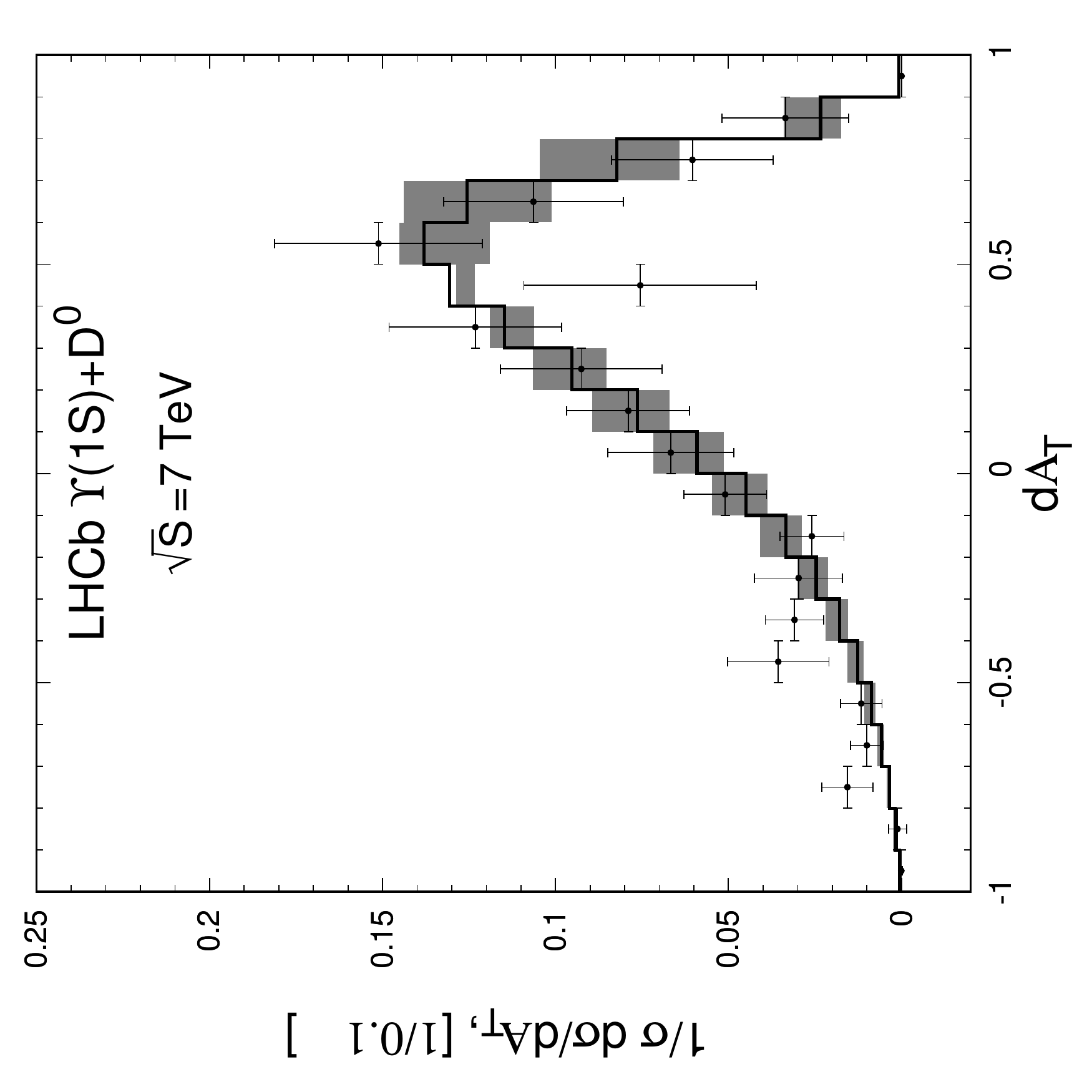}\includegraphics[width=0.5\textwidth, angle=-90,origin=c, clip=]{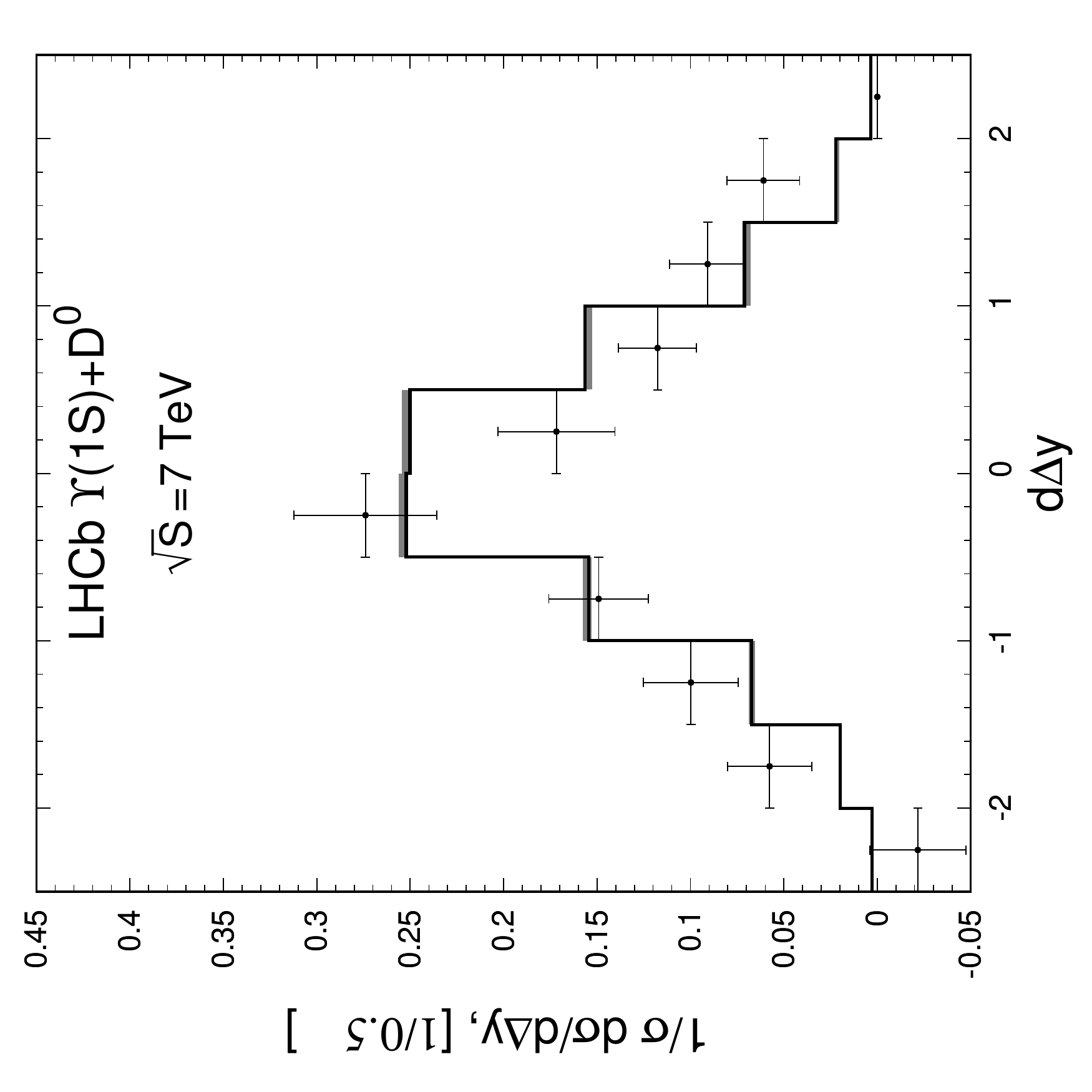}
\includegraphics[width=0.5\textwidth, angle=-90,origin=c, clip=]{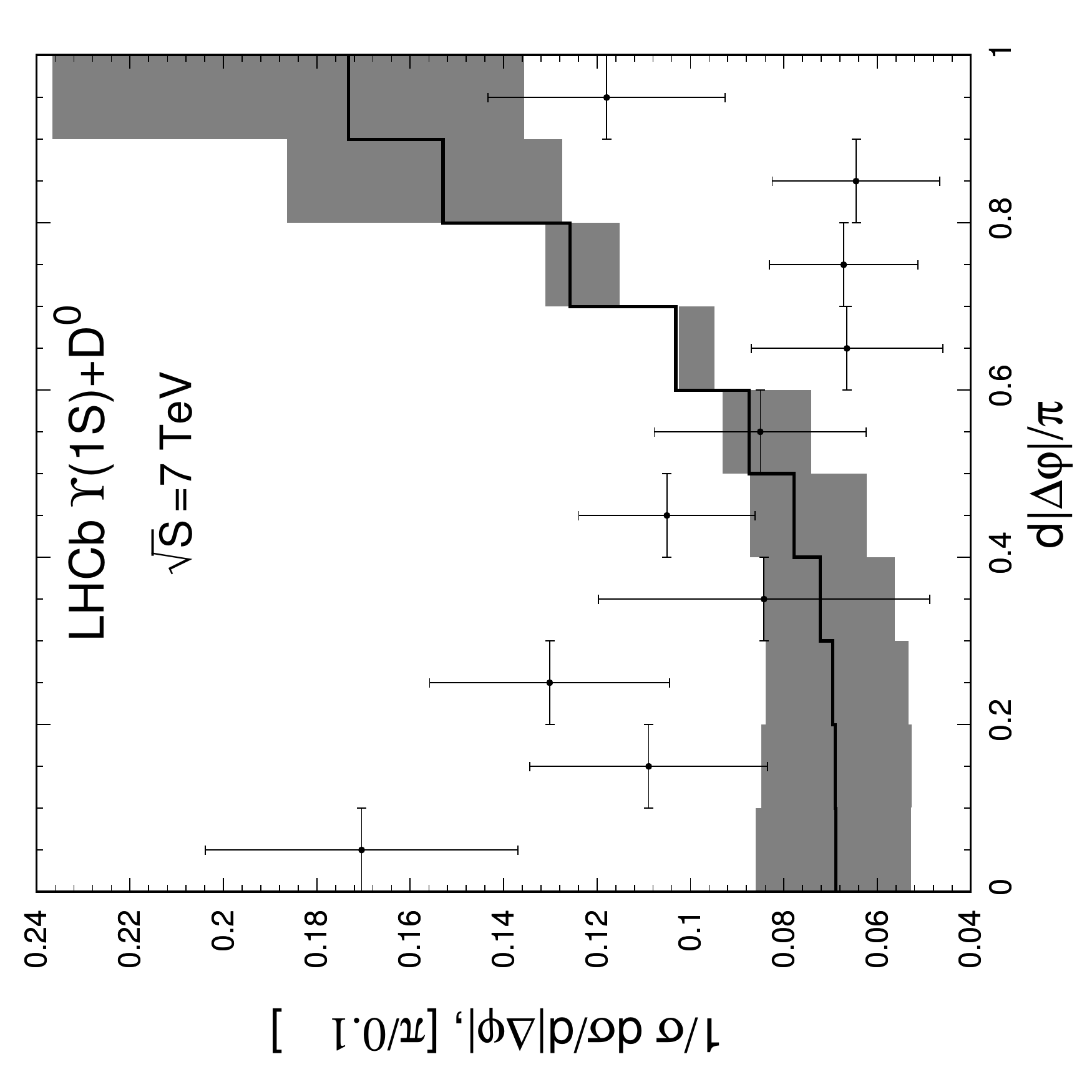}\includegraphics[width=0.5\textwidth, angle=-90,origin=c, clip=]{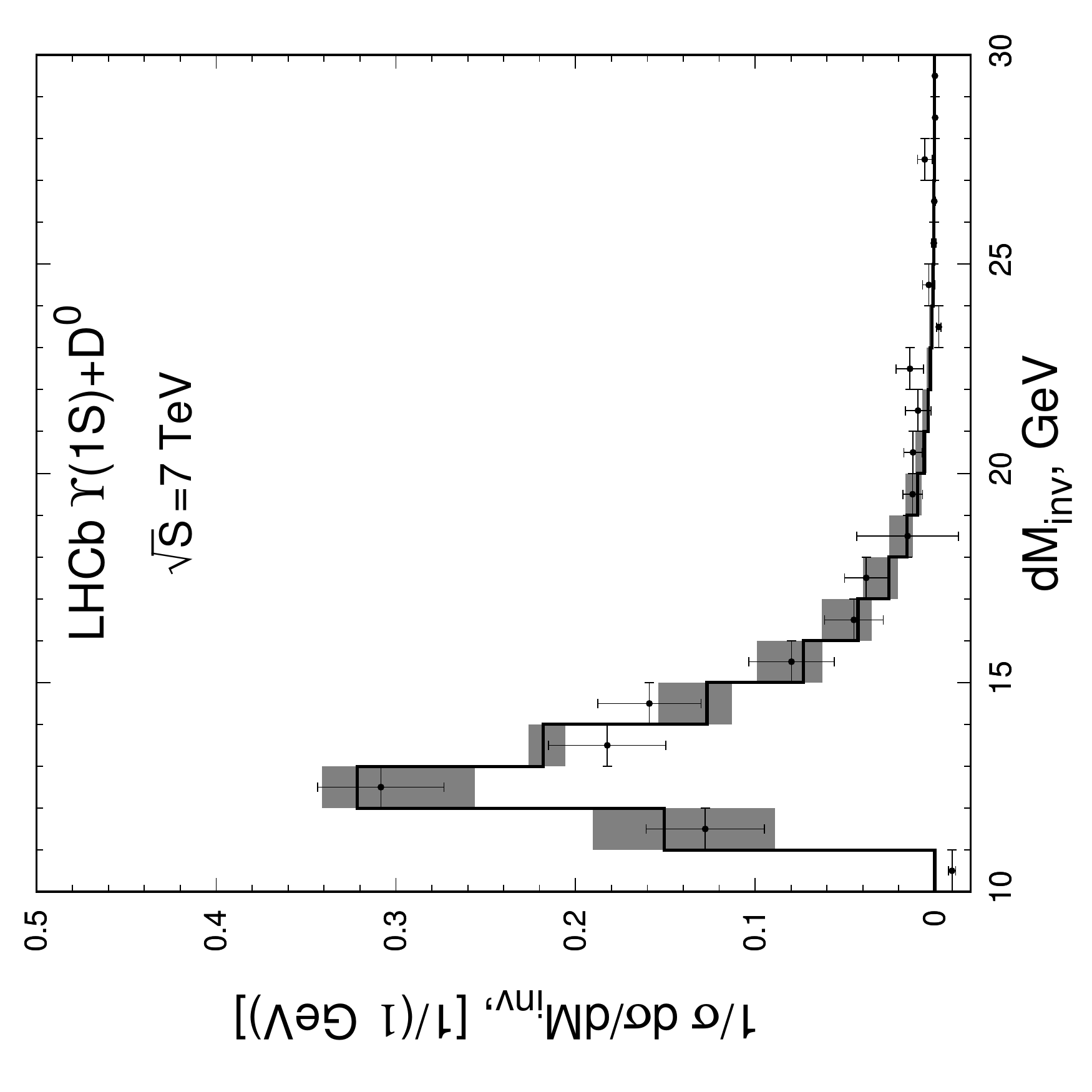}
 \caption{Spectra of  $\Upsilon(1S)+D^0$ associated production. The data from LHCb at the $\sqrt S=7$ TeV, $2.0<y_{\Upsilon(D)}<4.5$, $0<p_{T\Upsilon}< 15$ GeV, and $1<p_{TD}< 20$ GeV.}
 \label{fig-3}
\end{figure}

In the Fig. \ref{fig-3}, we plot spectra for the following correlation
variables: $M_{inv}$ is the invariant mass of $\Upsilon D$ pair,
$\triangle y=|y_{\Upsilon}-y_{D}|$ is the rapidity difference,
$\triangle \phi=|\phi_{\Upsilon}-\phi_{D}|$ is the azimuthal angle
difference, $A_T=(p_{T\Upsilon}-p_{TD})/(p_{T\Upsilon}+p_{TD})$ is
the transverse momentum asymmetry. Theoretical predictions for
$M_{inv}$, $\triangle y$ and $A_T$ spectra agree well with
experimental data. The $\triangle\phi$ spectrum in LO of
PRA has typical shape for this kind of spectra. It has one peak at the $\triangle\phi\sim \pi$ and
plateau at the $\triangle\phi\leq \pi/2$. However, the experimental
data from LHCb Collaboration, though having large errors,
demonstrate existence of the second peak near the value $\triangle\phi\simeq
0$. This feature of the data remains unexplained.

Below we estimate relative contributions of different production
mechanisms in the calculated total cross section $\left({\cal
B}_{\Upsilon(1S)\to\mu^+\mu^-}\times \sigma^{\Upsilon D^0}\right)$,
which is about 50 pb. Feeddown contribution from decays of higher
bottomonium states $(2S, 3S, 2P)$ is small, about 20 \%. In the
direct $\Upsilon(1S) D^0$ channel, the {\it gluon to $D-$meson}
fragmentation (\ref{RRg}) dominates over $c-$quark fragmentation
(\ref{RRcc}), 40 pb and 5 pb, correspondingly. There are
color-singlet ($b\bar{b}\left[^3S_1^{(1)}\right]$) and color-octet
($b\bar{b}\left[^3S_1^{(8)}\right]$) contributions in the the
subprocess (\ref{RRg}), the last one is dominant and gives about 30
pb. Having in mind that uncertainty of our calculation from choice
of hard scale and color-octet matrix elements is about factor 2, we
found that SPS contribution, calculated in the PRA, can account for
approximately one half of experimental measured value of
cross-section $\left({\cal B}_{\mu^+\mu^-}\times \sigma_{\rm
exp.}^{\Upsilon D^0}\right)=155\pm 21\pm 7$ pb. Such a way, contrary
to the conclusion of Ref. \cite{LHCb_ups_D}, the contribution of DPS
production mechanism is not dominant in the associated
$\Upsilon(1S)D$ production.

\section*{Acknowledgments}
Authors thank the Ministry of Education and Science of the Russian
Federation for financial support in the framework of the Samara
University Competitiveness Improvement Program among the world's
leading research and educational centers for 2013-2020, the task
number 3.5093.2017/8.9. The work of A.K. and M.N. is supported in
part by the Russian Foundation for Basic Research through the Grant
No. MK 18-32-00060. A.K. and M.N. thank the II Institute for
Theoretical Physics of Hamburg University for hospitality and
personally, Prof. B.A. Kniehl and Dr. Zhi-Guo He for stimulating
discussions.

\end{document}